\documentclass[preprint,prd,nofootinbib,preprintnumbers]{revtex4-2}
\usepackage{amssymb,amsmath,bm, mathrsfs, color}
\usepackage{amsfonts}
\usepackage{amstext}
\usepackage{graphicx}
\usepackage{color}
\usepackage[normalem]{ulem}
\usepackage{slashed}
\usepackage{hyperref}
\usepackage{textcomp}

\definecolor{mygray}{gray}{0.6}
\newcommand{\mygray}{\textcolor{mygray}}
\newcommand{\bea}{\begin{eqnarray}}
\newcommand{\eea}{\end{eqnarray}}

\begin{document}

\title{\Large \textbf{ Pseudo-Goldstone Dark Matter in $SO(10)$}}

\author{\large Nobuchika Okada\footnote{\color{blue} {okadan@ua.edu}}, Digesh Raut\footnote{\color{blue} {draut@udel.edu}}, Qaisar Shafi\footnote{\color{blue} {qshafi@udel.edu}}, and Anil Thapa\footnote{\color{blue} {thapaa@okstate.edu}} }

\affiliation{$^a$Department of Physics and Astronomy,University of Alabama, Tuscaloosa, Alabama 35487, USA}
\affiliation{$^{b,c}$ Bartol Research Institute, Department of Physics and Astronomy,University of Delaware, Newark DE 19716, USA}
\affiliation{$^d$ Department of Physics, Oklahoma State University, Stillwater, OK 74078, USA}

\begin{abstract}
We propose a pseudo-Goldstone boson dark matter (pGDM) particle in $SO(10)$ grand unified theory (GUT).  
Due to its Goldstone nature, this pGDM evades the direct DM detection experiments 
which, otherwise, severely constrain the parameter space of DM models.  
In $SO(10)$, the pGDM is embedded as a linear combination of the Standard Model (SM) singlet scalars in ${\bf 16_H}$ and ${\bf \overline{126}_H}$ representations.  
We consider two scenarios for the intermediate route of $SO(10)$ symmetry breaking (SB) to the SM: 
  $SU(5) \times U(1)_X$ and Pati-Salam the $SU(4)_c \times SU(2)_L \times SU(2)_R$ (4-2-2) gauge groups. 
The vacuum expectation value of ${\bf \overline{126}_H}$, which triggers the breaking of  $U(1)_X$ and 4-2-2 symmetry
  in the two scenarios, respectively, determines the pGDM lifetime
  whose astrophysical lower bound provides one of the most stringent constraints. 
For the 4-2-2 route to $SO(10)$, the successful SM gauge coupling unification requires the 4-2-2 breaking scale
  to be ${\cal O} (10^{11})$ GeV, and most of the parameter space is excluded. 
For the $SU(5) \times U(1)_X$ route, on the other hand, the $U(1)_X$ breaking scale can be significantly higher,  
  and a wide range of the parameter space is allowed. 
Furthermore, the proton lifetime in the $SU(5)$ case is predicted to be $4.53 \times 10^{34}$ years, 
  which lies well within the sensitivity reach of the Hyper-Kamiokande experiment. 
We also examine the constraints on the model parameter space from the Large Hadron Collider 
  and the indirect DM search by Fermi-LAT and MAGIC experiments.  
\end{abstract}

\maketitle

{\hypersetup{linkcolor=black}}
\setcounter{footnote}{0}
\newpage
\section{Introduction}\label{SEC-01}

Astrophysical and cosmological observations \cite{Lisanti:2016jxe} have provided compelling evidence for the existence of dark matter (DM), presumably a non-baryonic particle which constitutes about 85\% 
  of the observed matter energy density in the universe \cite{Aghanim:2018eyx}. 
A (quasi-)stable, electrically neutral weakly interacting massive particle (WIMP), absent in the Standard Model (SM), 
  is arguably one of the most attractive DM candidates. 
The interaction cross-section of WIMPs with the SM particles is typically subject to stringent bounds
  from the direct DM detection experiments for a wide range of the DM mass.

In Ref.~\cite{Gross:2017dan} the authors have proposed a SM Higgs-portal DM scenario 
  based on a softly broken global $U(1)$ symmetry, which is called ``pseudo-Goldstone DM (pGDM)".
The model includes a SM singlet complex scalar field $S$ with a $U(1)$ global charge, 
  whose vacuum expectation value (VEV) spontaneously breaks the $U(1)$ symmetry, 
  and a soft $U(1)$ symmetry breaking term, 
\bea
  \mu_S^2 \left( S^2 + (S^\dagger)^2 \right),  
 \label{eq:eq:pGDMglobal}  
\eea
where $\mu_S$ is a real mass parameter. 
The resulting scalar potential is invariant under $S\to S^\dagger$, so that  the model has a ${\bf Z}_2$ symmetry 
  under which only the imaginary component of $S$ (called $\chi$) has an odd-parity, 
  while the real component of $S$ and all the SM fields have even-parity.
Hence, the pseudo-Goldstone scalar $\chi$, which is the Goldstone boson associated 
  with the spontaneous $U(1)$ symmetry breaking in the limit of $\mu_s \to 0$, 
  is a stable DM candidate. 
Thanks to its Goldstone nature, the coupling of the pGDM $\chi$ with the SM Higgs boson
  is proportional to its momentum. 
As a result, the scattering cross-section of the DM particle with a nucleon through the Higgs boson exchange
  vanishes in the non-relativistic limit, thus evading the direct DM detection constraints \cite{Gross:2017dan}. 

An ultraviolet (UV) completion of this pGDM scenario was proposed in Refs.~\cite{Abe:2020iph, Okada:2020zxo}. 
The model is based on the gauged $B-L$ (Baryon number minus Lepton number) extended SM \cite{Pati:1974yy,Davidson:1978pm,Mohapatra:1980qe,Marshak:1979fm, Wetterich:1981bx,Masiero:1982fi,Mohapatra:1982xz,Buchmuller:1991ce}, 
  in which the accidental global $B-L$ symmetry of the SM is gauged. 
In this model, the soft breaking term of Eq.~(\ref{eq:eq:pGDMglobal}), which is crucial to realize the pGDM scenario, 
  is obtained from 
\bea
  \Lambda \left( \Phi_A \Phi_B^2 + {\rm h.c} \right),   
 \label{eq:pGDMBL}  
\eea
where $\Phi_{A,B}$ are complex scalars with $B-L$ charges $Q_A$ and $Q_B = -Q_A/2$, 
  and $\Lambda$ is a mass parameter.  
The fields $\Phi_{A,B}$ both develop VEVs and break the $B-L$ symmetry. 
The pGDM is realized as a linear combination of the imaginary components of $\Phi_{A,B}$.
If the imaginary component of $\Phi_B$ dominates, we approximately reproduce Eq.~(\ref{eq:eq:pGDMglobal}) 
  by replacing $\Phi_A \to \langle \Phi_A \rangle$ and identifying 
  $\Lambda \langle \Phi_A \rangle$ and $\Phi_B$ with $\mu_S^2$ and $S$, respectively. 
This is a key idea for the UV completion of the pGDM scenario.   
It is important to note, however, that the $B-L$ gauge interaction explicitly breaks the ${\bf Z}_2$ symmetry
  responsible for the stability of the pGDM, and hence the pGDM can decay via the $B-L$ gauge interaction. 
To satisfy the lifetime bound for a decaying DM, $\tau_{DM}> 10^{26}$ sec, from cosmic ray observations \cite{Essig:2013goa}, 
  the $B-L$ symmetry breaking scale is required to be $v_{BL} \gtrsim 10^{11}$ GeV \cite{Okada:2020zxo}. 
In addition, since the Goldstone nature of the DM particle is lost due to the cubic scalar coupling of Eq.~(\ref{eq:pGDMBL}), 
  the direct DM detection cross section does not exactly vanish  \cite{Okada:2020zxo}.

In this paper we propose a UV completion of the pGDM scenario in an $SO(10)$ grand unified theory (GUT), or in other words, an embedding of the gauge extension of pGDM in $SO(10)$.  
The $SO(10)$ grand unification is an interesting paradigm for the unification of strong, weak, and electromagnetic forces
  into a single force at a high energy scale. 
Each family of the SM quarks and leptons belongs in a single irreducible ${\bf 16}$ representation of $SO(10)$
  along with a SM singlet fermion. 
Electric charge quantization is incorporated, and the SM singlet fermion in the ${\bf 16}$ representation is identified with the right-handed neutrino, 
  which paves a natural way for tiny neutrino masses generation
  by the seesaw mechanism \cite{Minkowski:1977sc, Yanagida:1979as, Glashow:1979nm, GellMann:1980vs, Mohapatra:1979ia}. 
We will show in the following how the pGDM scenario can be successfully implemented in the $SO(10)$ framework.

The basic idea is as follows.
We introduce a scalar field in the ${\bf 16}$ representation (${\bf 16_H}$)\footnote{
While the ${\bf 16_H}$ must acquire a VEV in the pGDM scenario, 
  various $SO(10)$ DM models for a ${\bf 16_H}$ scalar with a vanishing VEV
  are discussed in \cite{Kadastik:2009dj,Kadastik:2009cu,Ferrari:2018rey}.}.  
Along with a Higgs field in the ${\bf 126}$ representation (${\bf \overline{126}_H}$), 
  whose VEV generates Majorana mass terms for the right-handed neutrinos, 
  consider the following $SO(10)$ invariant term in the scalar potential: 
\bea
\Lambda  {\bf \overline {126}_H} ({\bf 16_H})^2+ {\rm h.c}.  
\eea
Comparing this to Eq.~(\ref{eq:pGDMBL}), one can identify $\Phi_A$ and $\Phi_B$ with the SM singlet scalars 
  contained in ${\bf \overline{126}_H}$ and ${\bf 16_H}$, respectively. 
In the following,  we will consider two symmetry breaking patterns commonly used to obtain the SM from $SO(10)$ 
  via one intermediate step, namely, $SU(4)_c \times SU(2)_L \times SU(2)_R$ and $SU(5) \times U(1)_X$. 
The high VEV scale of ${\bf \overline{126}_H}$, which is necessary for the phenomenological viability of the (gauged) pGDM scenario, 
  can be naturally realized in SO(10) unification. 
We show a correlation between pGDM physics and GUT phenomenology, 
   such as a successful gauge coupling unification and observable proton decay. 
We carry out detailed phenomenological studies for the two symmetry breaking patters, 
   taking into account the constraints from the Large Hadron Collider and the indirect DM detection experiments, in order
   to identify the allowed parameter space.


\section{Phenomenology of PGDM}\label{SEC-02} 
Although our goal is to realize the pGDM scenario in the context of $SO(10)$, 
  the $B-L$ gauge extended model of the pGDM scenario contains all the essential features. 
For the pGDM phenomenology, it is sufficient to consider the following scalar potential involving $\Phi_{A,B}$ 
  and the SM Higgs doublet ($H$):  
\begin{align}
    V_{{\rm eff}} &= - \mu_H^2\ H^\dagger H - \mu_A^2\ \Phi_A^\dagger \Phi_A - \mu_B^2\ \Phi_B^\dagger \Phi_B + \lambda_H (H^\dagger H )^2 + \lambda_{HA} (H^\dagger H) (\Phi_A^\dagger \Phi_A) \nonumber \\
    &+ \lambda_{HB} (H^\dagger H) (\Phi_B^\dagger \Phi_B) + \lambda_{AB} (\Phi_A^\dagger \Phi_A) (\Phi_B^\dagger \Phi_B) + \lambda_A (\Phi_A^\dagger \Phi_A)^2 + \lambda_B (\Phi_B^\dagger \Phi_B)^2 \nonumber \\
    &- \{ \Lambda\ \Phi_A \Phi_B^2 + {\rm H.c.} \}. 
    \label{eq:Veff}
\end{align}
We represent the Higgs fields as follows:
\begin{equation}
    H = \begin{pmatrix}
         G^+   \\
         \frac{(h + v_H + i G^0)}{\sqrt{2}}
        \end{pmatrix} \, , \hspace{10 mm} 
        \Phi_{A, B} = \frac{1}{\sqrt{2}} (\phi_{A,B} + v_{A,B} + i \chi_{A, B}),
        \label{eq:Hspec}
\end{equation}
where $v_{H,A,B}$ are the VEVs of $H$, $\Phi_A$, and $\Phi_B$ fields, resepctively, with $v_H = 246$ GeV. 
Requiring the scalar potential to exhibit a minimum around its VEV location leads to the following relations: 
\begin{align}
    \mu_H^2 &= \frac{\lambda_{HA}\ v_A^2}{2} + \frac{\lambda_{HB}\ v_B^2}{2} + \lambda_{H}\ v_H^2 \, , \nonumber \\
    \mu_A^2 &= \lambda_A v_A^2 + \frac{\lambda_{HA}\ v_H^2}{2} + \frac{\lambda_{AB}\ v_B^2}{2} - \frac{v_B^2 \Lambda}{\sqrt{2}\ v_A} \, , \nonumber \\
    \mu_B^2 &= \lambda_B v_B^2 + \frac{\lambda_{AB}\ v_A^2}{2} + \frac{\lambda_{HB}\ v_H^2}{2} - \sqrt{2}\ v_A \Lambda \, .
\end{align}
After spontaneous symmetry breaking the Goldstone modes $G^{\pm,0}$, are absorbed by the SM gauge bosons $W^\pm$ and $Z$.
The remaining mixing mass matrices are expressed as 
\bea
V  \supset 
\frac{1}{2}
\begin{bmatrix}
\chi_A & \chi_B
\end{bmatrix}.{\cal M}_{{\rm I}}^2.
\begin{bmatrix} 
\chi_A \\ \chi_B 
\end{bmatrix} + 
\frac{1}{2} \begin{bmatrix}
h & \phi_B & \phi_A
\end{bmatrix}.{\cal M}_{{\rm R}}^2. \begin{bmatrix} 
h \\ \phi_B \\ \phi_A 
\end{bmatrix}, 
\label{eq:massmatrixABh}
\eea
where the mass matrices for the $CP$-odd scalars ($\chi_{A}, \chi_{B}$) and $CP$-even  scalars ($\phi_{A},\phi_{B}$, $h$)
  are given by 
\begin{equation}
    M^2_{{\rm I}} =
    \sqrt{2}\Lambda \begin{pmatrix}
               \frac{v_B^2}{2 v_A} & v_B \\
                 v_B & 2 v_A  \\
             \end{pmatrix}, 
    M_{{\rm R}}^2 = 
    \begin{pmatrix}
     2 \lambda_H v_H^2  & \lambda_{HB} v_B v_H & \lambda_{HA} v_A v_H \\
    \lambda_{HB} v_B v_H & 2 \lambda_{B} v_B^2  & v_B (\lambda_{AB} v_A - \sqrt{2} \Lambda) \\
    \lambda_{HA} v_A v_H & v_B (\lambda_{AB} v_A - \sqrt{2} \Lambda) & 2 \lambda_{A} v_A^2 + \frac{v_B^2 \Lambda}{\sqrt{2}  v_A}
    \end{pmatrix}
    \label{eq:massMat}
\end{equation}

In the $CP$-odd sector, one combination of $\chi_A$ and $\chi_B$ is the would-be Nambu-Goldstone mode ($\tilde{G}$) 
   absorbed by the $B-L$ gauge boson and the orthogonal combination is the pGDM $\chi$: 
\begin{equation}
    {\tilde G} = -\cos\theta \;\chi_A + \sin\theta \; \chi_B, 
    \hspace{10 mm} 
     \chi = \sin\theta \;\chi_A + \cos\theta\; \chi_B, 
    \label{eq:GX}
\end{equation}
where $\cos\theta = 2 v_A /\sqrt{4 v_A^2 + v_B^2}$, $\sin\theta = v_B /\sqrt{4 v_A^2 + v_B^2}$, and the mass of DM $\chi$ is given by 
\begin{equation}
    m_\chi^2 = 2 \sqrt{2}\Lambda v_A \left(1+\tan^2\theta\right). 
    \label{eq:massX}
\end{equation}
For the $CP$-even sector, we consider the limiting case for the parameters in the mass matrix, namely, $v_A \gg v_B, v_H$, $\lambda_A v_A^2 \gg \Lambda v_B$, and $\lambda_{HA}, \lambda_{AB} \to 0$. 
In this limit, $\phi_A$ decouples from the low energy effective theory, and 
  we only consider the sub-matrix for $h$ and $\phi_B$. 
This sub-matrix is diagonalized for the mass eigenstate, $\tilde{h}$ and $\tilde{\phi}_{B}$, 
  defined as \begin{align}
    \tilde{\phi}_B &= \cos\theta_H\ \phi_B + \sin\theta_H\ h \, , \\
    \tilde{h} &= - \sin\theta_H\ \phi_B + \cos\theta_H\ h \, .
\end{align}
Their mass eigenvalues are given by  
\begin{equation}
    m_{ \tilde{B}, \tilde{h}}^2 = \frac{1}{2} \bigg( m_B^2 + m_h^2 \pm \frac{m_B^2 - m_h^2 }{\cos 2\theta_H}\bigg) \, ,
    \label{eq:mass}
\end{equation}
where $m_B^2 = 2 \lambda_B v_B^2$, $m_h^2 = 2 \lambda_H v_H^2$, and the mixing angle $\theta_H$ is given by
\begin{equation}
    \tan 2\theta_H = \frac{2 v_B v_H \lambda_{HB}}{m_{B}^2 - m_{h}^2} \, .
\end{equation}


\subsection{Direct pGDM Detection Amplitude}
Let us calculate the direct DM detection amplitude for the elastic scattering of DM with a nucleon. 
The scattering occurs at very low energies, so zero momentum transfer limit, $t \to 0$\footnote{For $t\neq0$ the maximum value of the DM direct detection cross section is approximately $\theta_H^2 t_{\rm max}^2 \times \sigma_{HP}$ \cite{Abe:2020iph}, where $\sigma_{HP}$ is the Higgs-portal DM cross section, $\theta_H \simeq 0.1$, $|t_{\rm max}| = m_{DM}^2 v^2$ and $v\simeq 10^{-3}$ is the DM velocity. Hence, realistically we can expect the DM direct detection cross section to be closer to the neutrino floor. See, for example, \cite{Arcadi:2019lka}.}, is a good approximation to evaluate the scattering amplitude. 
In the limiting case with $v_A\gg v_B$, the pGDM $\chi \equiv \chi_B$ to leading order in $\theta$, such that the amplitude in the interaction basis $(h, \phi_B, \phi_A)$ is expressed as
\bea
\mathcal{M} &\propto& \begin{pmatrix}
  \frac{1}{2} \lambda_{HB} v_H \hspace{3mm} &  \lambda_B v_B \hspace{3mm}& \frac{1}{2} \lambda_{AB} v_A + \frac{1}{\sqrt{2}} \Lambda
\end{pmatrix} \,
(M_R^2)^{-1} \,
\begin{pmatrix}
Y_{hf\overline{f}} \\
0\\
0
\end{pmatrix}
+ {\cal O} (\theta^2) 
\nonumber \\
&\simeq& -\frac{\Lambda  \lambda_{HB}}{\sqrt{2} \lambda_{AB} \lambda_H v_A v_H} + {\cal O} \bigg(\frac{v_B^2}{4v_A^2}\bigg), 
\label{eq:zeroDDD}
\eea
where $Y_{hf\overline{f}}$ is the interaction of $h$ with the SM fermions.
Using Eq.~(\ref{eq:massX}), we obtain 
\bea
|{\cal M}|^2 &\propto& \frac{m_\chi^4}{v_A^4 v_H^2}. 
\label{eq:zeroDD}
\eea
It is clear that the direct DM detection amplitude is negligibly small.

\section{DM relic density and Indirect Detection}\label{SEC-04}
The pGDM can interact with the SM particles through the $\tilde h/\tilde{\phi}_B$ portal interaction as follows:   
\begin{equation}
    \mathcal{L} \supset \frac{\lambda_{\tilde{h}}}{2}\ \tilde{h} \chi \chi + \frac{\lambda_{\tilde{B}}}{2}\ \tilde{\phi}_B \chi \chi \, , 
\end{equation}
with
\begin{align}
    \lambda_{\tilde{h}} = - \frac{m_{\tilde{h}}^2}{v_B}\ \sin\theta_H \, , \hspace{20mm}
     \lambda_{\tilde{B}} =  \frac{m_{\tilde{B}}^2}{v_B}\ \cos\theta_H \, .
\end{align}
The couplings of $\tilde{h}$/$\tilde{\phi}_B$ to the SM fermions are as follows: 
\begin{equation}
    \mathcal{L} \supset -\ (\tilde{h} \cos\theta_H + \tilde{\phi}_B \sin\theta_H) \sum_f \frac{m_f}{v_H} \Bar{f} f.
\end{equation}

\begin{figure}[t!]
    \centering
    \includegraphics[height=6.1cm, width=8.15cm]{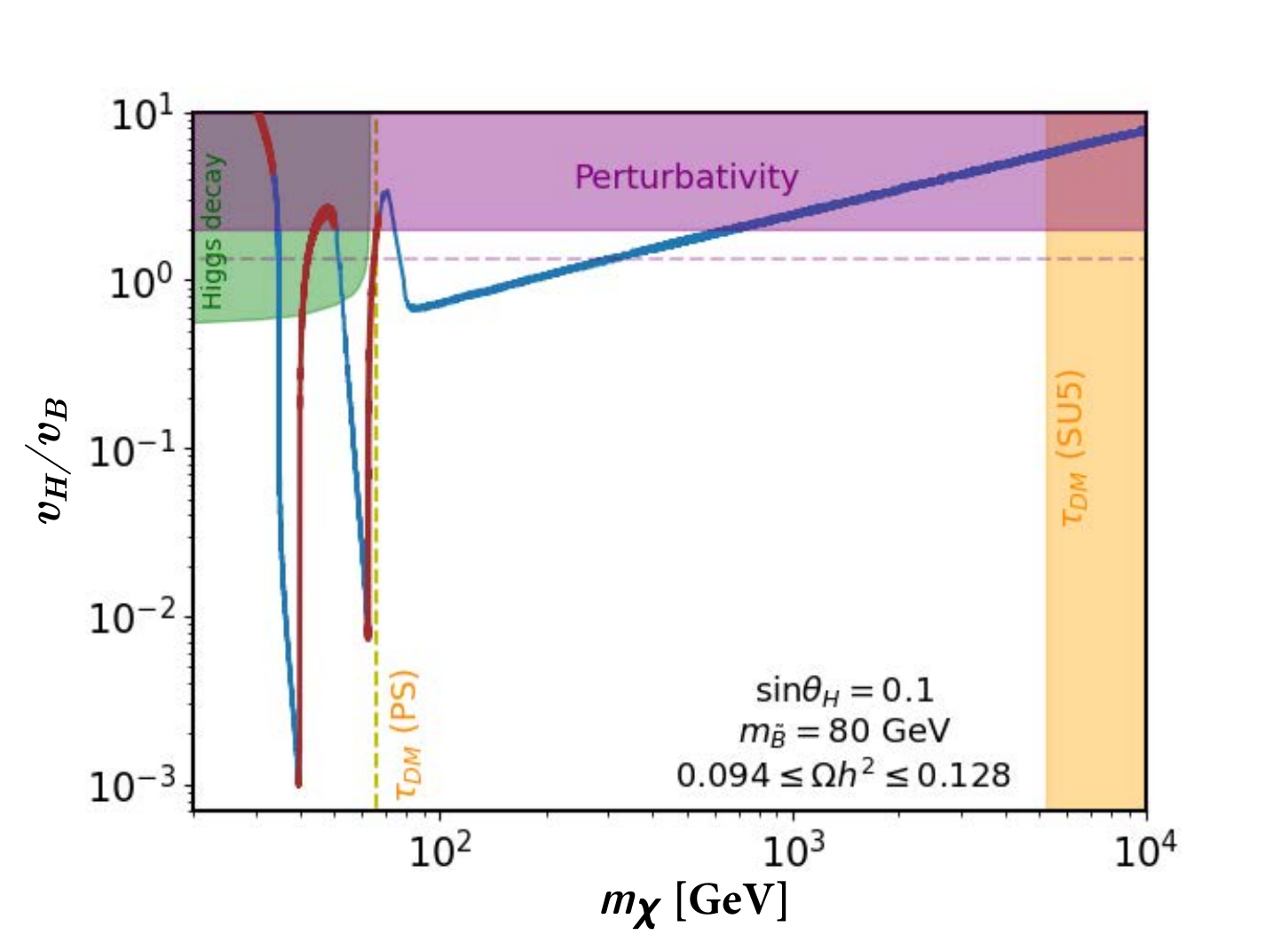}
    \includegraphics[height=6.1cm, width=8.15cm]{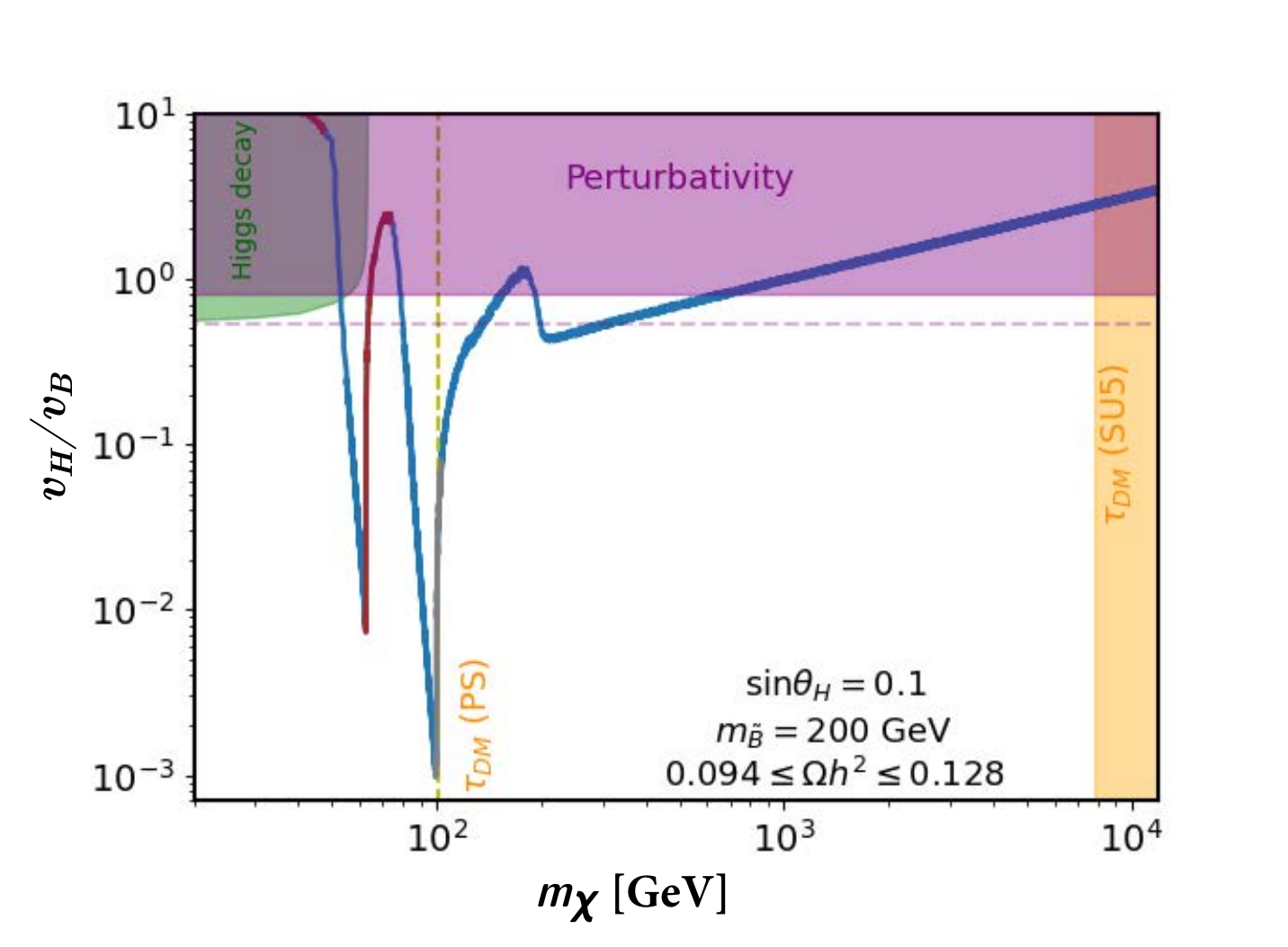}
    \includegraphics[height=6.1cm, width=8.25cm]{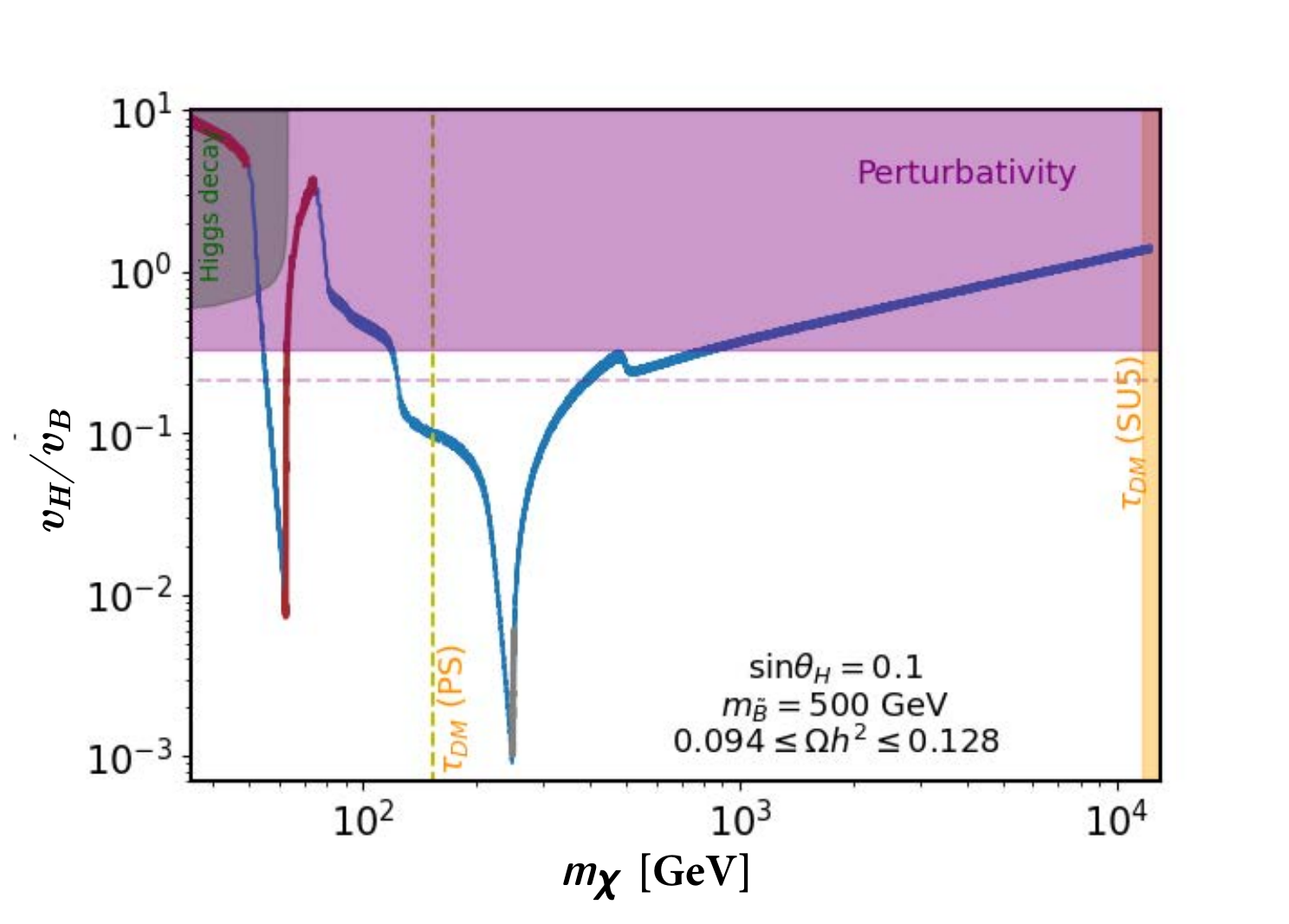}
    \caption{The allowed region in the pGDM mass $m_{\chi}$ and VEV ratio $v_H/v_B$ plane 
    for three different benchmark values $m_{\tilde B}/{\rm GeV} = 80, 200, 500$, and $\sin\theta_H = 0.1$ for the mixing between $h$ and $\phi_B$. The thermal relic abundance of $0.094 \leq \Omega h^2 \leq 0.128$ \cite{Aghanim:2018eyx} is satisfied along the blue curves. The green shaded regions are excluded by the Higgs invisible decay constraint \cite{Sirunyan:2018owy}. The brown and gray shaded segments along the each abundance curve is excluded after imposing {\tt Fermi-LAT  + MAGIC} bound obtained from annihilation of pGDM particles to $b\bar{b}$ and $W^+W^-$ (see Fig.~\ref{fig:indirect}). 
For the $SU(5)\times U(1)_X$ case, the parameter region depicted by the orange and the purple band respectively are excluded by the pGDM lifetime bound, $\tau_{DM} \gtrsim 10^{26} $ sec \cite{Essig:2013goa}, and perturbativity bound, namely, for $\lambda_B$ to remain perturbative up to the $SO(10)$ unification scale. Similarly, the orange and purple dashed line respectively depict the boundary of the excluded region for the 4-2-2 case.}
    \label{fig:relic}
\end{figure}

We numerically evaluate the relic abundance of the DM using the software {\tt MicrOmegas}  \cite{Belanger:2014vza}. The model is implemented in {\tt CalcHEP} \cite{Belyaev:2012qa} by using {\tt LanHEP} \cite{Semenov:2014rea}. 
The DM relic abundance depends on the pGDM pair annihilation cross section, which is determined by 4 free parameters,  $m_{\chi}$, $\theta_H$, $v_B$ and $m_{\tilde{B}}$. 
In the following analysis, we fix $\sin\theta_H =0.1$ as our benchmark and impose the criterion that the DM relic abundance $\Omega h^2 = 0.120\pm0.001$ \cite{Aghanim:2018eyx} measured by {\tt PLANCK} collaboration at 3$\sigma$ confidence level is satisfied.

A pair of DM particles annihilating into the SM particles which subsequently decay to produce gamma-rays can be searched by indirect DM detection experiments such as Fermi-LAT \cite{Ackermann:2015zua} and MAGIC \cite{Aleksic:2013xea}. 
These experiments have set an upper bound on the cross section of DM pair annihilation into $b\bar{b}$ 
and $W^+W^-$ final states. 
If $m_{\chi}< m_{\tilde h}/2$, the SM(-like) Higgs boson can decay into a pair of pGDM particles. 
The CMS result provides an upper bound on the branching ratio of the invisible Higgs boson decay, $BR( {\tilde h} \to \chi \chi) \leq 0.16$ \cite{Sirunyan:2018owy}. 
As previously mentioned, the pGDM gauge interaction explicitly violates the ${\bf Z}_2$ symmetry, which would otherwise forbid the decay of the pGDM. 
Hence we also impose a lifetime bound on pGDM, $\tau_{DM}> 10^{26}$ sec \cite{Essig:2013goa}, to further constrain the parameter space. 
In the following, we show the results of our analysis, including the lifetime constraints for both 4-2-2 and $SU(5) \times U(1)_X$ scenarios, leaving out the details of the pGDM decay, which will be presented in the later sections.

Our overall conclusions is that the allowed region for the 4-2-2 scenario is significantly smaller than for the $SU(5) \times U(1)_X$ case. 
This is because the VEV of ${\bf \overline{126}_H}$ ($v_A$) in the 4-2-2 case required by the successful SM gauge coupling unification
is at the intermediate scale, $10^{11}$ GeV, which is much smaller than the VEV scale required in the $SU(5) \times U(1)_X$ case. 
In particular, for the later case, the pGDM can be as heavy as $m_\chi/{\rm GeV} = 690, 715,860$ for $m_{\tilde B}/{\rm GeV} = 80, 200,500$, respectively. 
In contrast, only the region close to the resonance points survive for the 4-2-2 case with the pGDM mass $m_\chi \gtrsim {\cal O} (10^2)$ GeV excluded.    

\begin{figure}[t!]
    \centering
    \includegraphics[height=6cm, width=8cm]{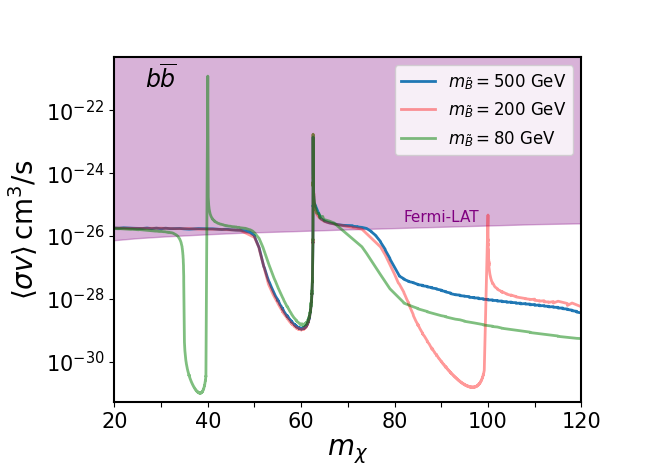}
    \includegraphics[height=6.1cm, width=8.3cm]{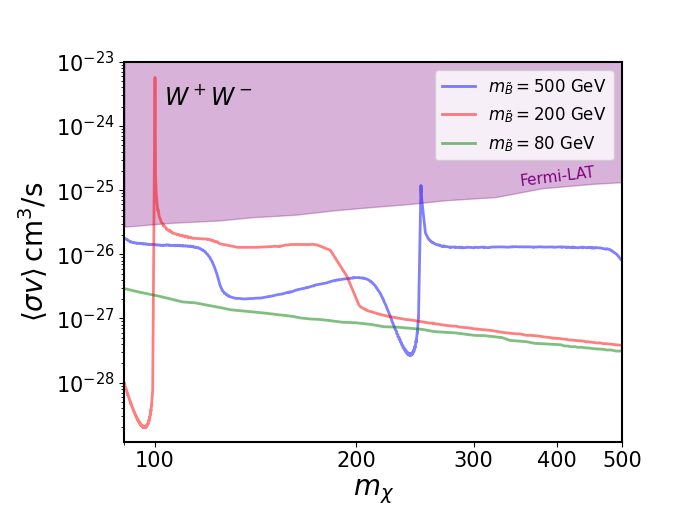}
    \caption{The left and right panels respectively show the DM annihilation cross section to $b\bar{b}$ and $W^+W^-$ for various values of the pGDM mass along with the exclusion (purple band) from {\tt Fermi-LAT + MAGIC} experiments \cite{Ahnen:2016qkx} for various benchmark values of $m_{\tilde B}$.}
    \label{fig:indirect}
\end{figure}

In Fig.~\ref{fig:relic}, we show the allowed region consistent with observed relic abundance as a function of the DM mass $m_{\chi}$ 
  and the ratio of the VEVs $v_H/v_B$ for three different choices of $m_{\tilde B}= 80, 200,500$ GeV. 
Here, the mixing angle between $h$ and $\phi_B$ is taken to be $\sin\theta_H =0.1$. 
The thermal relic abundance $0.094 \leq \Omega h^2 \leq 0.128$ \cite{Aghanim:2018eyx} is satisfied along the blue curve. 
The brown and gray shaded line segments on top of the blue curve are excluded by imposing 
the {\tt Fermi-LAT + MAGIC}  bound \cite{Ahnen:2016qkx} on the DM pair annihilation cross section 
  into $b\bar{b}$ and $W^+W^-$ (see Fig.~\ref{fig:indirect}). 
The green region is excluded by the constraint on the Higgs invisible decays,  $BR(\tilde{h} \to \chi \chi) < 0.16$, 
  obtained from the CMS result \cite{Sirunyan:2018owy}.
The orange band depicts the parameter region excluded in the $SU(5) \times U(1)_X$ case, whereas the region to the right of the dashed orange line is excluded for the 4-2-2 case in order to satisfy the pGDM lifetime bound from cosmic ray observations, 
 $\tau_{DM} \gtrsim 10^{26}$ sec \cite{Essig:2013goa}.  


The value of $\lambda_B$ is determined at low energy for fixed values of $m_{\tilde B}$,  $v_B$ and $\theta_H$ in Eq.~(\ref{eq:mass}). 
Using this value, we have evaluated the renormalization group running of $\lambda_B$ up to the $SO(10)$ unification scale ($v_{GUT}$). 
Requiring $\lambda_B$ to remain perturbative at $v_{GUT}$, we obtain an upper bound on $v_H/v_B$ (see Sec.~\ref{SEC-05} and \ref{SEC-06} for details). 
For the $SU(5)\times U(1)_X$ case, the excluded region is depicted by the purple band, whereas the region above the dashed purple line is excluded for the 4-2-2 case.

In Fig.~\ref{fig:indirect}, we show the DM annihilation cross section to $b\bar{b}$ (left) and $W^+W^-$ (right) 
  for various values of DM mass along with the upper bounds on the cross section 
  from {\tt Fermi-LAT + MAGIC} experiments \cite{Ahnen:2016qkx}. 
The pGDM pair annihilation cross sections are evaluated by using $v_B$ and $m_{DM}$ values
  that satisfy the relic abundance constraint in Fig.~\ref{fig:relic}. 
The green, red, and blue curves show the results for $m_{\tilde B}/{\rm GeV} = 80, 300$, and $500$, respectively. 
The purple band represents the exclusion from {\tt Fermi-LAT + MAGIC} experiments \cite{Ahnen:2016qkx}. 

\section{$SO(10)$ Breaking via $SU(5) \times U(1)_X$}\label{SEC-03}

A simple way to embed the pGDM scenario in $SO(10)$ GUT is based on the generalization
  of the minimal ${B-L}$ scenario to the so-called gauged $U(1)_X$ extended SM \cite{Oda:2015gna}.
The generalized $U(1)_X$ charges ($Q_X$) of each particles are defined 
 to be a linear combination of its $B-L$ charge ($Q_{BL}$) and hypercharge ($Q_Y$), $Q_X = x_H Q_Y + Q_{BL}$, 
 where $x_H$ is a free parameter. 
The interesting feature of the $U(1)_X$ extended SM is that for $x_H = -4/5$, the SM quarks and lepton
 are unified in ${\bf 10}$ and ${\bf {\overline 5}}$ representations of $SU(5)$ with 
 $U(1)_X$ charges $+1/5$ and $-3/5$, respectively  \cite{Okada:2017dqs}. 
In the $SO(10)$ embedding, the $U(1)_X$ charge is normalized to be $Q_X \to \sqrt{5/8}\; Q_X$\cite{Okada:2020vvb}.

The Higgs fields involved in breaking  $SO(10)$ down to the SM along with their decomposition under $SU(5) \times U(1)_X$ are listed below: 
\begin{align}
    {\bf 45_H} &= {\bf 1_H}_{0}  \oplus {\bf 24_H}_{0} \mygray{ \oplus {\bf 10_H}_{-\sqrt{2/5}} \oplus {\bf \overline{10_H}}_{\sqrt{2/5}}}\, ,\nonumber \\
   {\bf  {126}_H} &= {\bf 1_H}_{\sqrt{5/2}} \mygray{\oplus {\bf5_H}_{\sqrt{1/10}} \oplus {\bf \overline{10_H}}_{\sqrt{3/10}} \oplus {\bf15_H}_{-\sqrt{3/10}} \oplus {\bf\overline{45_H}}_{-\sqrt{1/10}} \oplus {\bf50_H}_{\sqrt{1/10}}} \, , \nonumber\\
   {\bf  16_H}  &= {\bf 1_H}_{\sqrt{5/8}} \mygray{\oplus  {\bf\overline{5_H}}_{-3/{\sqrt 40}} \oplus  {\bf10_H}_{1/{\sqrt 40}}} \nonumber \, ,\\
    {\bf 10_H}  &=  {\bf5_H}_{-1/{\sqrt 10}} \mygray{\oplus  {\bf\overline{5}}_{1/{\sqrt 10}}} \, .
    \label{eq:SU5dec}
\end{align}
Here only the $SU(5) \times U(1)_X$ multiplets depicted in bold develop nonzero VEVs to trigger the following symmetry breaking chain: 
\begin{align}
S O(10)\ \xrightarrow[v_{GUT}]{ \langle {\bf 45_H} \rangle}\ SM \times U(1)_X \xrightarrow[v_A,v_B]{\langle {\bf \overline{126}_H} \rangle, \langle {\bf16_H} \rangle} \ SM  \xrightarrow[v_{EW}]{\langle {\bf10_H} \rangle} SU(3)_c \times U(1)_{EM}, 
\label{eq:PSSB}
\end{align}
Here, $v_{GUT}$, $v_{A,B}$, and $v_{EW}$ denote the VEVs of the corresponding scalars fields involved in the various stages of the symmetry breaking.

The Higgs potential relevant for the discussion of pGDM at low energies is expressed as  
\begin{align}
    V & \supset -\mu_1^2 ({\bf10_H^\dagger 10_H}) - \mu_2^2 ({\bf\overline{126}_H^\dagger \overline{126}_H}) - \mu_3^2 ({\bf16_H^\dagger 16_H})  + \lambda_1 ({\bf10_H^\dagger 10_H})^2+\lambda_2 ({\bf16_H^\dagger 16_H})^2  \nonumber \\
    &+\lambda_3 ({\bf\overline{126}_H^\dagger \overline{126}_H})^2 + \lambda_4 ({\bf10_H^\dagger 10_H})({\bf\overline{126}_H^\dagger \overline{126}_H}) +  \lambda_5\ ({\bf\overline{126}_H^\dagger \overline{126}_H})({\bf16_H^\dagger 16_H})  
    \nonumber  \\
    &+\lambda_6\ ({\bf10_H^\dagger 10_H})({\bf16_H^\dagger 16_H})- \big({\Lambda_1 {\bf \overline{126}_H} ({\bf16_H})^2 + {\rm h.c}}\big). 
    \label{eq:potSU5}
\end{align}
Here we have excluded the $SO(10)$ invariant terms ${\bf 10_H (16_H)}^2$ and ${\bf 10_H^\dagger (16_H)}^2$,
 since they explicitly violate the Goldstone nature \cite{Gross:2017dan} by allowing a direct coupling between pGDM and SM Higgs $H\supset {\bf 10_H}$.
The SM singlet components in ${\bf \overline{126}_H}$ and ${\bf 16_H}$ are identified to be $\Phi_A$ and $\Phi_B$, and the SM Higgs doublet is contained in the ${\bf 5}$-plet of ${\bf 10_H}$. 
With a suitable choice of $\mu_{2,3}$ to balance the contribution of other scalars to the ${\bf 16_H}$ and ${\bf \overline{126}_H}$ masses, the $\Phi_{B}$ and $\Phi_{A}$ masses can be arranged to be at the TeV scale and GUT scale, respectively.  
The electroweak scale mass for the SM Higgs $H$ can be generated by the well known triplet-doublet splitting mechanism using the ${\bf 10_H}$ coupling with the $SU(5)$ adjoint ${\bf 24_H} \supset {\bf 45_H}$. 
Hence the low energy effective scalar potential matches the pGDM potential in Eq.~(\ref{eq:Veff}).

It is known that the unification of the three SM gauge couplings can be achieved 
  by introducing new vector-like quark/lepton pairs at certain mass scales \cite{Amaldi:1991zx,Chkareuli:1994ng,Choudhury:2001hs,Morrissey:2003sc,Gogoladze:2010in,Chen:2017rpn,Okada:2017dqs,Okada:2018tgy,Okada:2020cvq}. 
Let us introduce a new vector-like fermion pair in ${\bf 16}+{\bf \overline{16}}$ representation under $SO(10)$ 
  and consider the following Lagrangian \cite{Okada:2020vvb}: 
\bea
{\cal L} &\supset& {\bf {\overline {\bf 16}}} \big( Y {\bf 45_H} - M\big) {\bf 16} \nonumber \\
&\supset& 
{\bf {\overline 5}}\big(a_5 Y \langle {\bf 1_H}\rangle + b_5 Y \langle {\bf 24_H}\rangle - M \big){\bf 5}
+ {\bf {\overline {10}}}\big(a_{10} Y \langle{\bf 1_H}\rangle + b_{10} Y \langle {\bf 24_H}\rangle -  M \big){\bf 10}, 
\eea
where $Y$ is a Yukawa coupling constant, $M$ is a Dirac mass, and 
  the second line denotes the decomposition under the $SU(5)$ sub-group 
  with $a_i$ and $b_i$ being the Clebsch-Gordan coefficients. 
Assuming $\langle {\bf 1_H} \rangle={\cal O}(v_{GUT})$, 
  $\langle {\bf 24_H} \rangle = {\cal O}(v_{GUT}) {\rm diag} (1,1,1,-3/2,-3/2)$ and $M={\cal O}(v_{GUT})$
  for the symmetry breaking of $SO(10) \to SM \times U(1)_X$, 
  we obtain the mass splitting among the vector-like fermions in different SM representations. 
Tuning the Yukawa coupling value of $Y$ allows us to make one vector-like fermion pair light 
  while the others have mass at the GUT scale. 
We introduce three vector-like fermion pairs of  ${\bf 16}+{\bf \overline{16}}$ representation 
   and tune three Yukawa couplings to provide three vector-like fermion pairs below the GUT scale, 
   namely, one new down-type quark pair $(D^c+ \overline{D^c})$ with mass $M_{D^c}$, 
   one new quark-doublet pair $(Q+\overline{Q})$ with mass $M_Q$  
   and one new lepton-doublet pair $(L + \bar{L})$ with mass $M_L$.

\begin{figure}[t]
\begin{center}
\includegraphics[scale=0.7]{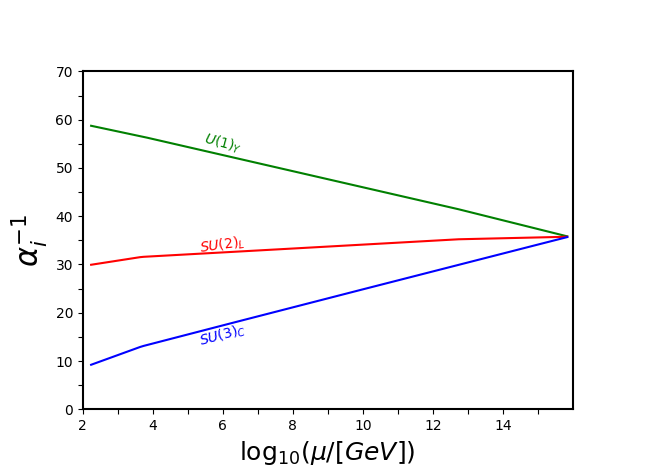}
\end{center}
\caption{
RG running of the SM ($\alpha_{1,2,3}$) and $U(1)_X$ ($\alpha_{X}$) gauge couplings. 
}
\label{fig:gcuSU5}
\end{figure}

For the energy scale $\mu < v_{GUT}$, 
  the renormlaization group (RG) equations of the SM and the $U(1)_X$ gauge couplings
  at the 1-loop order are given by 
\bea
 \mu  \frac{d {\alpha}_{1}}{d \mu} &=& \frac{1}{2\pi}{\alpha}_{1}^2 \left(\frac{41}{10} + \frac{2}{5}\times \rm{\theta(\mu -M_Q)}+  \frac{2}{5} \times \rm{\theta(\mu -M_L)}\right)
 \nonumber \\
 \mu  \frac{d {\alpha}_{2}}{d \mu} &=& \frac{1}{2\pi}{\alpha}_{2}^2 \left(-\frac{19}{6} + 2\times \rm{\theta(\mu -M_Q)}+  \frac{2}{3} \times \rm{\theta(\mu -M_L)}\right), 
 \nonumber \\
 \mu  \frac{d {\alpha}_{3}}{d \mu} &=& \frac{1}{2\pi}{\alpha}_{3}^2
\left(-7 + 2\times \rm{\theta(\mu -M_Q)} \right),
 \nonumber \\
 \mu  \frac{d {\alpha}_{X}}{d \mu} &=& \frac{1}{2\pi}{\alpha}_{X}^2
\left(\frac{169}{60}  + \frac{11}{10}\times \rm{\theta(\mu -M_Q)} + \frac{3}{5}\times \rm{\theta(\mu -M_L)} \right),
 \label{eq:RGSU51}
\eea
Here $\alpha_{i} \equiv g_{i}^2/4\pi$, $g_{2,3}$ respectively are the $SU(3)_c$ and $SU(2)_L$ SM gauge couplings, $\alpha_1 = g_1^2/4\pi$ is related to the SM hypercharge gauge coupling by $g_{1} = \sqrt{5/3}\; g_Y$, and 
we have set $M_{D^c}=M_Q$ for simplicity. 
In our RG analysis, we employ the low energy values of the SM gauge couplings at $\mu = m_t = 172.44$ GeV \cite{Buttazzo:2013uya}: 
\bea
g_{1} (m_t) = \sqrt{5/3} \times 0.35830, \qquad
g_{2} (m_t) =  0.64779 , \qquad 
g_{3} (m_t) = 1.1666.  
\eea 
We find that the $SU(5)$ gauge coupling unification condition, 
   $\alpha_{1,2,3}(v_{GUT})=\alpha_{GUT}$, 
   is satisfied for benchmark values $M_{Q} = 4.8 \times 10^3$ GeV and $M_L = 4.5\times 10^{12}$ GeV, 
   and we obtain  $v_{GUT} = 7.58 \times 10^{15}$ GeV with $1/\alpha_{GUT} = 35.8$. 
The $SO(10)$ unification requires $\alpha_X(v_{GUT})=\alpha_{GUT}$. 
The RG running of the SM and $U(1)_X$ gauge couplings are shown in Fig.~\ref{fig:gcuSU5}.
Using this result, the proton lifetime mediated by the $SO(10)$ GUT gauge bosons can be approximated as \cite{Nath:2006ut}
\bea
\tau_p \approx \frac{1}{\alpha_{GUT}^2} \frac{v_{GUT}^4}{m_p^5} \approx 9.61 \times 10^{34} \; {\rm years}, 
\label{eq:PD}
\eea  
where $m_p = 0.983$ GeV is the proton mass. 
This is consistent with the current bound on proton lifetime obtained by the Super-Kamiokande experiment $\tau_p(p \to \pi^0 e^+) \gtrsim 1.6\times 10^{34}$ yr \cite{Miura:2016krn}. 
Importantly, this is within the expected sensitivity reach of the future Hyper-Kamiokande experiment, 
  $\tau_p \simeq1.3 \times 10^{35}$ yr \cite{Abe:2011ts}.
Since the RG running of the three SM couplings only depends on two free parameters, there is a one-to-one  correspondence between the masses $M_Q$ and $M_L$ to satisfy the unification of the couplings. 
As a result, we find that $ M_{Q} \gtrsim 3.4 \times 10^3$ GeV, or equivalently $ M_{L} \lesssim 8.0 \times 10^{12}$ GeV,  for the proton lifetime to be within the search reach of Hyper-Kamiokande experiment.

Next let us discuss the gauge interaction of the pGDM which explicitly breaks the ${\bf Z}_2$-parity and thereby enabling it to decay.  
For instance, the gauge interaction in the $B-L$ extended model \cite{Okada:2020zxo} is given by
\bea
{\cal L} \supset -g_{BL} \Big(Q_B \phi_B \cos\theta+ Q_A \phi_A \sin\theta \Big)  (\partial_\mu \chi) {Z_{BL}^\mu} , 
\label{eq:kinlag}
\eea
where $Q_{A,B}$ is the $B-L$ charge of $\Phi_{A,B}$. 
Because $\phi_A$ is much heavier than $\phi_B$ the dominant decay mode is $\chi \to {Z^\prime}^* \phi_B^* \to {\bar f}_{SM} f_{SM} {\bar f}_{SM} f_{SM}$. 
In the $SO(10)$ model, we can identify $Z_{BL} \to Z_X$ with $Q_A = -\sqrt{5/2}$ and $Q_B = \sqrt{5/8}$.  
For the interaction of the $Z_X$ gauge boson with the SM $SU(2)_L$ singlet quarks ($q$) and leptons ($\ell$)  we obtain  
\bea
{\cal L} \supset -g_X\Big( Q_q \; {\overline q} \gamma^\mu  q + Q_\ell\;  {\overline \ell} \gamma^\mu \ell \Big) {Z_{X\mu}},
\label{eq:ZXSM}
\eea
where $Q_q = -3/\sqrt{40}$ and $Q_\ell = +1/\sqrt{40}$. 
The $Z_X$ mass is given by
\bea
m_{Z_X} = \sqrt{\frac{5}{8}} g_X \sqrt{ 4 v_A^2 +  v_B^2}   \simeq \sqrt{\frac{5}{2}} g_X v_A, 
\eea 
where we have used $v_B\ll v_A$ in the last expression. Using Eqs.~(\ref{eq:kinlag}) and ~(\ref{eq:ZXSM}), we have evaluated the lifetime of the pGDM. 
We impose the bound on its lifetime $\tau_\chi>10^{26}$ sec from the cosmic ray observations \cite{Essig:2013goa} 
  to obtain a lower bound on the pGDM mass. 
The results for various benchmark values of $m_{\tilde B}/{\rm GeV} =  80, 200, 500$ are depicted
  by the yellow band in Fig.~\ref{fig:relic}. 
Here, we have fixed $v_A=v_{GUT}$.

For the RG equation of the $\Phi_B$ quartic coupling  $\lambda_B \equiv \lambda_1$ in Eq.~(\ref{eq:potSU5}) we obtain 
%
\bea 
 \mu  \frac{d {\lambda_B}}{d \mu} &=& \frac{1}{16\pi^2} 20\lambda_{B}^2. 
\eea
for $\mu < v_{GUT}$. 
We impose the perturbative bound on $20 \lambda_B^2/16\pi^2< 1$ at $\mu = v_{GUT}$, 
  which is extrapolated to the upper bound on $\lambda_B(v_B)$. 
Thus, we find the lower bound on $v_B$ by using $m_{\tilde B}^2 \simeq 2 \lambda_B(v_B) v_B^2$. 
%
For ${\tilde m_B}/{\rm GeV} = 80, 200,500$, we obtain an upper bound of $v_H/v_B \lesssim 2.0, 0.81, 0.33$ with $\lambda_B (v_{GUT}) = 0.21,0.22,0.22$, respectively. 
These are depicted by the dashed purple line in Fig.~\ref{fig:relic}.

\vspace{-10pt}

\section{$SO(10)$ Breaking via $SU(4)_c \times SU(2)_L\times SU(2)_R$ (4-2-2)}\label{SEC-05}
The Higgs fields involved in breaking $SO(10)$ down to the SM along with their decomposition under the 4-2-2 gauge group ( $SU(4)_c \times SU(2)_L\times SU(2)_R$) are listed below  
\bea
{\bf 210_H} &=& ({\bf 1},{\bf 1},{\bf 1}) \mygray{ \oplus ({\bf 15},{\bf 1},{\bf 1}) \oplus ({\bf 6}, {\bf 2},{\bf 2}) \oplus ({\bf 15},{\bf 3},{\bf 1}) \oplus ({\bf 15},{\bf 1},{\bf 3}) \oplus ({\bf 10},{\bf 2},{\bf 2}) \oplus (\overline{\bf 10},{\bf 2},{\bf 2})} , \nonumber \\
{ {\bf \overline{126}_H}} &=& ({\bf 10},{\bf 1},{\bf 3}) \oplus ({\bf 15}, {\bf 2}, {\bf 2}) \mygray{ \oplus  ({\overline {\bf 10}}, {\bf 3}, {\bf 1}) \oplus ({\bf 6},{\bf 1},{\bf 1})} , \nonumber \\
{\bf 45_H} &=& ({\bf 1},{\bf 1},{\bf 3}) \oplus ({\bf 15},{\bf1},{\bf1}) \mygray{\oplus ({\bf 1},{\bf 3},{\bf 1}) \oplus ({\bf 6},{\bf 2},{\bf 2})}, \nonumber \\
{\bf 16_H} &=& ({\bf 4},{\bf 1},{\bf 2}) \mygray{\oplus ({\overline {\bf 4}}, {\bf 2},{\bf 1})}, \nonumber \\
{\bf 10_H}&=& ({\bf 1},{\bf 2},{\bf 2}) \mygray{\oplus ({\bf 6},{\bf 1},{\bf 1})}. \label{eq:PSdecomp}
\eea 
Here only the 4-2-2 multiplets depicted in bold develop a nonzero VEV, which results in the following symmetry breaking pattern: 
\begin{align}
SO(10)\ \xrightarrow[v_{GUT}]{ \langle {\bf 210_H} \rangle}\ {\rm PS}\ \xrightarrow[v_A,v_A,v_B]{\langle {\bf \overline{126}_H} \rangle, \langle {\bf45_H} \rangle, \langle {\bf16_H} \rangle} \ SM  \xrightarrow[V_H]{\langle {\bf10_H} \rangle} SU(3)_c \times U(1)_{EM}, 
\label{eq:PSSBp}
\end{align}
where $v_{GUT}$, $v_{A,B}$, and $v_{EW}$ denote the VEVs of the corresponding scalars fields involved at the various stages of the symmetry breaking. 

In the fermion sector, the ${\bf 10_H}$ and ${\bf \overline{126}_H}$ Higgs field generate realistic fermion masses \cite{Holman:1982tb, Babu:1992ia, Bajc:2001fe, Fukuyama:2002ch, Goh:2003sy, Goh:2003hf, Bertolini:2004eq, Babu:2005ia, Aulakh:2005bd, Bajc:2005qe, Aulakh:2005mw, Bajc:2005zf, Bertolini:2006pe, Bajc:2008dc, Bertolini:2009qj, Joshipura:2011nn, DiLuzio:2011mda, Bertolini:2012im, Bertolini:2012az, Altarelli:2013aqa, Dueck:2013gca, Babu:2015bna, Aulakh:2006hs, Aulakh:2008sn, Fukuyama:2015kra, Babu:2018tfi, Babu:2018qca, Ohlsson:2019sja},
while the SM singlet component in ${\bf (\overline{10},1,3)}$ of ${\bf \overline{126}_H}$ generates intermediate scale Majorana masses for the right-handed neutrinos in the ${\bf 16}$-plet fermion.

There are a total of four Higgs doublets in $({\bf 1},{\bf 2},{\bf 2})$ of ${\bf 10_{H}}$ and $({\bf 15}, {\bf 2}, {\bf 2})$ of ${\overline {\bf 126_{H}}}$. 
We assume all of them develop nonzero VEV at the electroweak scale such that only one of the four linear combination of the doublets is light (doublet-doublet Higgs mass splitting), which is identified as the SM Higgs doublet $H$. 
The rest of the linear combinations are assumed to have masses of order $v_A$. 
The SM singlets components in $({\bf 10},{\bf 1},{\bf 3})$ of ${\overline {\bf 126_{H}}}$ and $({\bf 4},{\bf 1},{\bf 2})$ of ${\bf 16_H}$ are identified to be $\Phi_A$ and $\Phi_B$, respectively. 
The inclusion of ${\bf 45}_H$ is crucial to realize the electroweak scale mass for $\Phi_B$ while making sure that all of the remaining charged scalars masses in $({\bf 4},{\bf 1},{\bf 2})$ are of the order $v_A$. 
Without ${\bf 45}_H$, all charged scalars will have the same mass as $\Phi_B$, which is inconsistent with the null observation of any elementary light-charged scalar. 
To illustrate the mass splitting, consider the following coupling: 
\bea
V &\supset& {\bf 16_H}^\dagger {\bf 45_H} {\bf 16_H}\nonumber \\
&\supset&(a_3 V_T - a_{15} V_{15})(\Phi_A^\dagger \Phi_A) + (y_3 V_T + y_{15} V_{15})(\Phi_Y^\dagger \Phi_Y) + (C_3 V_T +C_{15} V_{15})(\Phi_C^\dagger \Phi_C).  
\eea
Here $\Phi_Y$($\Phi_C$) is the component that is only charged under hypercharge ($SU(3)_c$),  $a_i,y_i$ and $C_i$ are Clebsch-Gordan coefficients which are of the same order, and $V_T$ and $V_{15}$ are respectively the VEVs of $({\bf 1},{\bf 1},{\bf 3})$ and $({\bf 15},{\bf 1},{\bf 3})$ in ${\bf 45_H}$ which are of order $v_{A}$.
By appropriately fixing $V_T\simeq V_{15}$, $\Phi_B$ can be light while the others fields are heavy.  
The $\Phi_A \supset ({\bf 10},{\bf 1},{\bf 3})$ has mass compatible to the scale $v_A$. 
As discussed before, we do not include the terms ${\bf 10_H (16_H)}^2$ and ${\bf 10_H^\dagger (16_H)}^2$ which explicitly violate the Goldstone nature of the pGDM \cite{Gross:2017dan}. 
The resulting scalar potential matches that of the low energy effective potential of pGDM in Eq.~(\ref{eq:Veff}). 

Let us now examine the RG evolution of the SM and 4-2-2 gauge couplings by solving their RG equations at the 1-loop level. 
For this analysis, we fix the masses of $({\bf 1},{\bf 1},{\bf 3})$ and $({\bf 15},{\bf 1},{\bf 3})$ in ${\bf 45_H}$\footnote{If $({\bf 1},{\bf 1},{\bf 3})$ and $({\bf 15},{\bf 1},{\bf 3})$ masses are lighter, the scale $v_A$ can be reduced, which can significantly lower the upper bound on the pGDM mass in Fig.~\ref{fig:relic}.} and all the multiplets highlighted in gray in Eq.~(\ref{eq:PSdecomp}) have masses of order $v_{GUT}$.   
For energy scale $\mu$ below the 4-2-2 scale ($\mu < v_A$), the RG equations of the SM gauge couplings are given by  
\bea
 \mu  \frac{d \alpha_{1}}{d \mu} &=& \frac{\alpha_{1}^2}{2\pi}
\left(\frac{41}{10}\right), 
 \nonumber \\
 \mu  \frac{d \alpha_{2}}{d \mu} &=& \frac{\alpha_{2}^2}{2\pi} \left(-\frac{19}{6} \right), 
\nonumber \\
 \mu  \frac{d  \alpha_{3}}{d \mu} &=& \frac{\alpha_{3}^2}{2\pi} \left(-7\right). 
\label{eq:betafun1}   
\eea
Their low energy values are fixed to match the measured values at $\mu = m_t = 172.44$ GeV \cite{Buttazzo:2013uya}:  
\bea
g_{1} (m_t) = \sqrt{5/3} \times 0.35830, \qquad
g_{2} (m_t) =  0.64779 , \qquad 
g_{3} (m_t) = 1.1666.  
\eea 
For $v_A < \mu < v_{GUT}$, our theory is based on the 4-2-2 gauge group. 
The relationships between the SM and the 4-2-2 gauge couplings at $\mu = v_A$ are given by the tree-level matching conditions:   
\bea
 \alpha_{2} (v_A) = {\alpha}_{L} (v_A) ,
 \qquad \alpha_{3} (v_A) = {\alpha}_{4} (v_A) , 
 \qquad
 \alpha_{1}^{-1} (v_A) = \frac{3}{5} {\alpha}_{R}^{-1} (v_A) + \frac{2}{5}{\alpha}_{4}^{-1}(v_A), 
\eea
Here $\alpha_{4,L,R}$ are $SU(4){_c}$, $SU(2)_L$ and $SU(2)_R$ gauge couplings and their RG equations are given by  
\bea
\mu  \frac{d {\alpha}_{4}}{d \mu} &=& \frac{1}{2\pi}{\alpha}_{4}^2 
\left(+1\right), 
\nonumber \\
 \mu  \frac{d {\alpha}_{L}}{d \mu} &=& \frac{1}{2\pi}{\alpha}_{L}^2 \left(4\right), 
\nonumber \\
 \mu  \frac{d {\alpha}_{R}}{d \mu} &=& \frac{1}{2\pi}{\alpha}_{R}^2
\left(\frac{32}{3}\right). 
\label{eq:betafun2}   
\eea

\begin{figure}[t]
\begin{center}
\includegraphics[scale= 0.65]{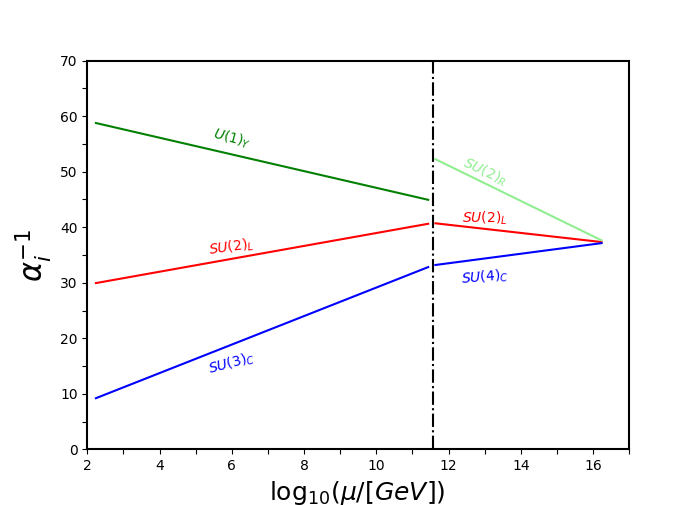} 
\end{center}
\caption{
The RG running of the SM gauge couplings ($\alpha_{1,2,3}$) for $\mu < v_A$ and 4-2-2 gauge couplings ($\alpha_{4,L,R}$) for $v_A< \mu < v_{GUT}$. 
}
\label{fig:gcuPS}
\end{figure}

With the initial values of the SM gauge couplings fixed at $\mu = m_t$, the RG equations for the SM and 4-2-2 gauge couplings can be solved analytically. 
After imposing the unification conditions 
\bea
\alpha_{4} (v_{GUT}) =\alpha_{L} (v_{GUT}) =\alpha_{R} (v_{GUT}) \equiv \alpha_{GUT}, 
\eea
$v_{GUT}$ and $v_{A}$ are uniquely determined. 
We find $v_A \simeq  3.6 \times 10^{11}$ GeV and $v_{GUT} \simeq 2.3 \times 10^{16}$ GeV, with $1/\alpha_{GUT} = 37.2$. 
Furthermore, using Eq.~(\ref{eq:PD}), the proton lifetime is predicted to be $8.8\times10^{36}$ yr. 
The RG running of the SM and 4-2-2 gauge  couplings are shown in Fig.~\ref{fig:gcuSU5}. 

Let us now estimate the lifetime of the pGDM. 
As discussed in the $SU(5) \times U(1)_X$ scenario, we are interested in the gauge interaction of pGDM $\chi$ with $\phi_B$ and the heavy neutral gauge boson $Z_R$ from the 4-2-2 symmetry breaking. 
This channel contributes dominantly to the pGDM decay rate. 
The heavy neutral gauge boson mass eigenstate is a linear combination of the gauge field associated with the  diagonal generators of $SU(4)_c$ and $SU(2)_R$ in the flavor basis, and its mass is given by 
\bea
m_{Z_R} = \frac{v_A}{2}\sqrt{3 g_4^2 + 2 g_R^2}  
\eea

\begin{figure}
    \centering
    \includegraphics[scale=0.6]{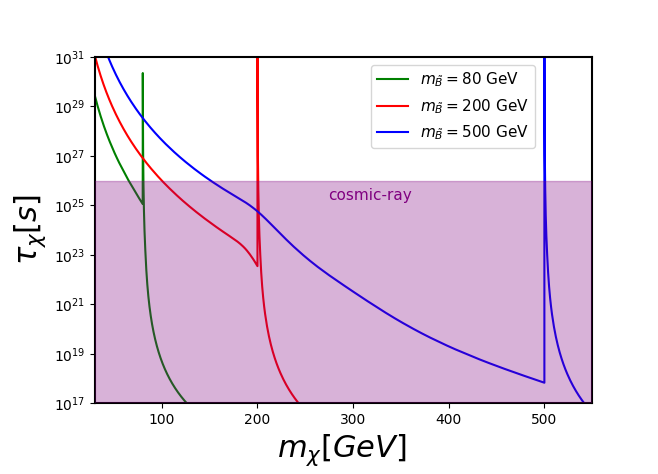}
    \caption{Lifetime of pGDM as a function of its mass ($m_\chi$). Green, red, and blue curves denote three benchmark scenarios with $m_{\tilde B}/{\rm GeV} =  80, 200, 500$, respectively. The purple band is excluded in order to satisfy the DM lifetime bound  $\tau_{\chi} \gtrsim 10^{26}$ from cosmic ray observations \cite{Essig:2013goa}. }
    \label{fig:4body}
\end{figure}

The relevant interactions for the pGDM decay, $\chi \to Z_{R}^* \phi_B^* \to \bar{f}_{SM} f_{SM} \bar{f}_{SM} f_{SM}$, are expressed as  
\bea
{\cal L} \supset  \Big(\frac{1}{2}  \kappa \phi_B \cos\theta - \kappa \phi_A \sin\theta \Big)  (\partial_\mu \chi) {Z_{R}^\mu} + \Big( Q_q \; {\overline q} \gamma^\mu  q + Q_\ell\;  {\overline \ell} \gamma^\mu \ell \Big) {Z_{R\mu}}, 
\label{eq:kinlag1}
\eea
with $\kappa = \sqrt{3 g_4^2 + 2 g_R^2}/2 = \sqrt{2} m_{Z_R}/v_A$, and the effective $Z_R$ charges for the $SU(2)_L$ singlet SM quarks ($q$) and $SU(2)_L$ doublet leptons ($\ell$) given by
\bea
Q_q &=& \frac{1}{12}\Big({\sqrt 6} g_4\sin\delta + 6 \cos\delta g_R \Big),\nonumber \\
Q_{\ell} &=& \frac{1}{4}\Big({\sqrt 6} g_4\sin\delta - 2 \cos\delta g_R \Big).
\eea
Here $\delta$ is the mixing between the $SU(4)_c$ and $SU(2)_R$ components of $Z_{R}$ defined by $\tan\delta = \sqrt{3/2} (g_4/g_R)$. 
In Fig.~\ref{fig:4body} we show the pGDM lifetime as a function of it mass and fixed mass of $\phi_B$ ($m_{\tilde B}$). 
The parameter region excluded by requiring  the pGDM lifetime $\tau_\chi>10^{26}$ sec, is depicted by  purple band. 
For $m_{\tilde B}/{\rm GeV} =  80, 200, 500$, the lifetime lower bound excludes the pGDM mass below $m_\chi/{\rm GeV} \simeq 66, 100, 153$, respectively. 
As shown in Fig.~\ref{fig:relic}, this significantly reduces the allowed parameter space for the pGDM due to the smallness of $v_A$ in the 4-2-2  case compared to the GUT scale $v_A$ value in the $SU(5)\times U(1)_X$ case. 

For the $\Phi_B$ quartic coupling $\lambda_B$ we obtain the following RG equation 
\bea 
 \mu  \frac{d {\lambda_B}}{d \mu} &=& \frac{1}{16\pi^2} \Big(20\lambda_{B}^2 + \theta(\mu -v_{A}) \Big(48 \lambda_B^2 + \frac{151875}{2048} g_4^4 + \frac{243}{128} g_R^4- \Big(\frac{675}{64} g_4^2 + \frac{27}{16} g_R^2\Big) \lambda_B\Big)\Big). 
\eea
Using the solution for the running of the 4-2-2 gauge couplings in Fig.~\ref{fig:gcuPS}, we can numerically evaluate the RG running of $\lambda_B$ together with  ${\tilde m_B}^2 \simeq 2 \lambda_B v_B^2 $ which fixes the initial value of $\lambda_B$ at $\mu = v_B$. 
For fixed values of ${\tilde m_B}$ and $v_B$ we impose the perturbativity bound $68 \lambda_B^2/16\pi^2< 1$ at $\mu = v_{GUT}$. 
For ${\tilde m_B}/{\rm GeV} = 80, 200,500$, we obtain an upper bound of $v_H/v_B \lesssim 1.3, 0.53, 0.22$ with $\lambda_B(v_{GUT})/10^{-2}  \simeq 9.3 ,9.4,9.5$, respectively. 
These are the depicted by dashed purple line in Fig.~\ref{fig:relic}.

\section{Intermediate and GUT Monopoles}\label{SEC-06}
Inflation can resolve the monopole problem in GUTs if the Hubble parameter during inflation ($H_{inf}$) is lower in magnitude than the monopole producing symmetry breaking scale. For instance, in $SO(10)$ breaking via 4-2-2, inflation driven by a gauge singlet with minimal coupling to gravity can, in principle, take care of both the GUT monopole as well as the intermediate mass monopole associated with the 4-2-2 breaking to the SM if the inflationary Hubble parameter is of order $10^{13} - 10^{14}$ GeV \cite{Shafi:2006cs}. A similar situation holds for quartic inflation driven by a gauge singlet field with non-minimal coupling to gravity \cite{Okada:2010jf, Okada:2014lxa}.
However, in our case with a 4-2-2 breaking scale significantly lower than $10^{13}$ GeV, it would be more appropriate to use the 4-2-2 breaking scalar field as the inflaton.
This scenario will be discussed in detail elsewhere \cite{Okada2021}.
Since there are only heavy GUT monpoles in the $SU(5) \times U(1)_X$ scenario, the non-minimal inflation by using a $SO(10)$ singlet field with Hubble parameter is of order $10^{13} - 10^{14}$ GeV \cite{Shafi:2006cs} nicely works to resolve the monopole problem.

\section{Conclusion}
The pGDM scenario is an interesting WIMP DM scenario which can avoid the very severe constraints from the direct DM search experiments. 
Because of the Goldstone nature of the pGDM, its coupling with the Higgs bosons vanishes in the limit of zero momentum transfer. 
In this paper, we have proposed an UV completion of the pGDM in the $SO(10)$ GUT framework. 
For this completion, it is crucial to introduce the $SO(10)$ invariant cubic scalar coupling ${\bf \overline{126}_H} {\bf 16_H}^2$ and the pGDM arises as a linear combination of the SM singlet scalars in ${\bf 16_H}$ and ${\bf \overline{126}_H}$.

We have considered two scenarios for the intermediate breaking patterns of $SO(10)$ to the SM:
$SU(5) \times U(1)$ and the 4-2-2 gauge group. 
Since the ${\bf 16_H}$ and ${\bf \overline{126}_H}$ Higgs fields are involved in breaking $SO(10)$ down to the SM, there is a direct correlation between dark matter and GUT physics. 
Specifically, the $SU(2)_R$ gauge boson in the 4-2-2 case and $U(1)_X$ gauge boson in the $SU(5) \times U(1)_X$ case allow pGDM to decay. 
Hence, to satisfy the cosmic ray bound on the pGDM lifetime, $\tau_{DM} \gtrsim 10^{26}$ s, we have shown that the vacuum expectation value of ${\bf \overline{126}_H}$, which triggers the 4-2-2 and $U(1)_X$ symmetry breaking, must be much greater than $10^{11}$ GeV in order to maximize the allowed parameter space for the pGDM. 
The 4-2-2 symmetry breaking scale occurs at the intermediate scale, $M_{PS} \simeq 2.3 \times 10^{11}$ GeV, which excludes most of the model parameter space. 
However, in the $SU(5) \times U(1)_X$ scenario, the $U(1)_X$ breaking scale can be much higher such that a  significantly larger portion of the model parameter space is still allowed. 
Furthermore, in the $SU(5)$ scenario, the proton lifetime is predicted to be $4.53 \times 10^{34}$ years, which lies well within the sensitivity range of the future Hyper-Kamiokande experiment. 
We have also examined the constraints on the model parameter space from the Large Hadron Collider and indirect DM searches by Fermi-LAT and MAGIC experiments.

\section*{Note Added}
During the write up of our manuscript we came across a recent paper \cite{Abe:2021byq} which discusses the same topic with $SO(10)$ broken via the gauge symmetry $SU(4)_c \times SU(2)_L \times SU2)_R$. Where there is overlap our results are in agreement. The $SO(10)$ symmetry breaking via $SU(5) \times U(1)_X$ has not been discussed in the above mentioned paper.

\section*{Acknowledgments}
This work is supported in part by the United States Department of Energy grant DE-SC0012447 (N. Okada), DE-SC0013880 (D. Raut and Q. Shafi), and DE-SC0016013 (A. Thapa).

\newpage
\bibliographystyle{utcaps_mod}
\bibliography{references}
\end{document}